# A flexible 300 mm integrated Si MOS platform for electron- and hole-spin qubits exploration

R. Li[1], N. I. Dumoulin Stuyck[1,2], S. Kubicek[1], J. Jussot[1], B. T. Chan[1], F. A. Mohiyaddin[1], A. Elsayed[1,3], M. Shehata[1,3], G. Simion[1], C. Godfrin[1], Y. Canvel[1], Ts. Ivanov[1], L. Goux[1], B. Govoreanu[1] & I. P. Radu[1]

[1]Imec, B-3001 Leuven, Belgium, email: roy.li@imec.be
[2]Department of Materials Engineering (MTM), KU Leuven, B-3001 Leuven, Belgium
[3]Department of Physics and Astronomy, KU Leuven, B-3001 Leuven, Belgium

*Abstract*—We report on a flexible 300 mm process that optimally combines optical and electron beam lithography to fabricate silicon spin qubits. It enables on-the-fly layout design modifications while allowing devices with either n- or p-type ohmic implants, a pitch smaller than 100 nm, and uniform critical dimensions down to 30 nm with a standard deviation ~ 1.6 nm. Various n- and p-type qubits are characterized in a dilution refrigerator at temperatures ~ 10 mK. Electrical measurements demonstrate well-defined quantum dots, tunable tunnel couplings, and coherent spin control, which are essential requirements for the implementation of a large-scale quantum processor.

## I. Introduction

Silicon quantum dot qubit systems are strong candidates for large-scale quantum processors. Long coherence times and high-fidelity operations have been demonstrated in fundamental qubit gates [1]. Although these devices are fabricated in lab environments, their structures are compatible with state-of-the-art silicon technologies. Upscaling a quantum processor crucially relies on the performance of each individual physical qubit [2]. However, a detailed understanding of the device and material properties at qubit operation temperature is still lacking. Efforts in scaling up silicon quantum processors beyond the few-qubit scale are limited by unpredicted parasitic effects.

In this article, we discuss our design cycles towards a better understanding of cryogenic material properties and present our latest developments on qubit fabrication and characterization. Fig. 1 depicts the concept of the design cycle, which is built upon the close link between fabrication, characterization, and device modelling. We develop a 300 mm process flow for qubit specific integration that produces high volume and uniform devices. Furthermore, it allows rapid design updates and enables the incorporation of complex structures, different gate materials, and either n- or p-type implants as ohmics. At 300 K, we perform standard semiconductor characterizations on both dedicated metrology structures and qubit structures for process control and for studying the influence of fabrication parameters. Cryogenic characterization consists of qubit measurements in a dilution refrigerator at temperatures in the milli-Kelvin range. We show controlled charge and spin operations, which can be correlated with process, design, and room temperature data. Qubits are simulated with a multi-physics model, which can be calibrated with cryogenic hardware data [3]. Finally, the updated model guides the design for the next cycle.

## II. Device Fabrication

The fabrication is based on a 300 mm process (Fig. 2(a)). Relaxed size features, including zero markers, ohmic junctions, spin-control stripline antennas or micro-magnets, and fanout metal are defined by optical lithography. The pitch-critical qubit gate structures are patterned by electron beam lithography (EBL) on 300 mm wafers in predefined qubit areas (Fig. 2(b)) [4].

We choose EBL here as a short turnaround replacement for advanced optical lithography, which would limit rapid device redesign and process. The EBL modules are fully co-integrated with optical lithography modules, in a single integrated process, which includes all needed blocks for qubit functionality. A specific back end of line (BEOL) passivation module and via contacting are developed, compatible with sample preparation requirements for cryogenic characterizations (Fig. 2(c-d)).

In the qubit region, we have the flexibility to incorporate different structures across a 300 mm wafer, from a double quantum dot device to a one-dimensional (1D) array (Fig. 3). We adopt a 3-layer overlapping gate structure with a pitch of 100 nm or below, allowing tight confinement potential around the quantum dots. Furthermore, the thickness of each gate metal is 20 ~ 30 nm and the isolation dielectric is ~ 5 nm. This reduces the topography influence so that each gate layer has similar coupling to the quantum dots underneath (Fig. 4). The geometry of the gates greatly affects the qubit operations and interactions. The width and pitch of the EBL gates is monitored by critical dimension scanning electron microscope (CDSEM) across the wafer. As an example, we monitor the confinement gate gap that defines the lateral size of all quantum dot in a 1D array (Fig. 5). We record the dimensions of the resist after EBL and gate after etching because they are important design parameters. Their 3σ are ~ 5 nm, confirming high EBL pattern uniformity. To summarize, the established fab recipes and inline monitoring guarantee uniformity and reproducibility, while the EBL qubit gates allow fast design turnaround. The whole integration makes systematic design variations possible, which is crucial for the following study of material properties and design criteria.

### III. 300 K CHARACTERIZATION

The 300 mm fab wafers are characterized with automated probe station systems. We incorporate different metrology structures across the wafer to extract material properties and monitor process steps. Fig. 6(a) shows the $I_dV_g$ studies of metrology transistors with different channel lengths on different gate layers. Long channel transistors show a systematic upward shift in the threshold voltage with higher gate layers, which have thicker dielectrics. However, this trend is not clear for short channel transistors due to short channel effects. Nonetheless, the narrow cumulative distribution function (CDF) of the threshold voltage signifies high process uniformity (Fig. 6(b)).

Qubits are tested at room temperature with customized probe cards (Fig. 7(a)). For EBL gate patterning, we can reliably achieve 30 nm spacing without shorts (Fig. 7(d)). Smaller spacings can be achieved with EBL recipes for smaller pitches. We further study $I_dV_g$ on single quantum dot structures (Fig. 8). The right barrier (RB) has a slightly larger CDF span than that of the left barrier (LB) with the same design size. It hints at EBL proximity or etching loading effects as RB is closer to the fanout regions of the qubit gates. 300 K characterization allows us to extract material properties and design topography related peculiarities systematically. This can be further studied for correlations to cryogenic results [5].

### IV. CRYOGENIC CHARACTERIZATION

To verify qubit device designs and study qubit control related properties, we characterize various qubit structures in a dilution refrigerator with a base temperature lower than 10 mK. We study single charge control with a similar gate structure as Fig. 4(b). We use barrier gates LB and RB to isolate a quantum dot along the channel induced by the top gate ST (Fig. 9(a)). This structure is also called single electron transistor (SET) for nMOS (or single hole transistor (SHT) for pMOS). In the small bias condition (source-drain bias $V_{sd} < 1$ mV), only a single electron (hole) can pass through the dot each time because of Coulomb repulsion. This structure can be described by a classical constant interaction model, and the equivalent circuit is shown in Fig. 9(b) [6]. We measure the dot conductance while sweeping LB and RB. The ~ 45° high conductance lines in Fig. 9(c) and (d) confirm a quantum dot formed between LB and RB. Fig. 10(a) shows the Coulomb current oscillations in a SET, denoting consecutive addition of individual electrons to the dot. On sweeping $V_{sd}$, the Coulomb oscillations develop into Coulomb diamonds, and the regular diamond patterns signify a well-defined quantum dot (Fig. 10(b)) [6].

Like a single dot, a double dot can be induced by stacking more gates. We define a double dot system with a similar structure as Fig. 3(b), where the quantum dots are defined under plunger gates 1 and 2 (P1 and P2). The schematic cross-section and the circuit model are shown in Fig. 11(a) and (b), respectively. For both nMOS and pMOS double dot devices, we can demonstrate separate control of the double dot potential with P1 and P2, and highly tunable inter-dot tunnel coupling with the barrier gate B2 (Fig. 11 (c)) [5,6].

The previous measurements characterize quantum dots in the many charge regime. However, single spin qubits generally require single charges in the quantum dots. In Fig. 12, we show that the quantum dots can be emptied to the very last charge. We measure a pMOS double dot device similar to Fig. 4, and the schematic cross-section is shown in Fig. 12(a). We operate the SHT dot as a potentiometer to sense the charges in dot 1 and 2 (equivalent circuit shown in Fig. 12(b)) [6]. By sweeping P1 and P2 voltages, we can empty the holes in dot 1 and 2 to the very last (Fig. 12(c)). We correct the gate crosstalk by remapping P1 and P2 onto a virtual gate space, allowing separate control of the potential on dot 1 and 2 [7]. This is shown by the almost vertical and horizontal transition lines in Fig. 12(d). Furthermore, we demonstrate tunable inter-dot tunnel coupling in the single hole state (Fig. 13). This is required for robust two qubit gates [8].

We now access the spin qubit operations. We start with pMOS qubits, as the strong spin-orbit coupling (SOC) of holes allows electrical spin control without the need for extra structures such as stripline antennas or micromagnets [9]. We induce a double quantum dot system and tune it to weak tunnel coupling regime, similar to Fig. 11(c) bottom left. With increased $|V_{sd}|$, the conductance point expands to a bias triangle (Fig. 14). In some charge configurations, the bottom of the bias triangle could be missing, as shown in Fig. 14(a) right. This is due to Pauli spin blockade (PSB) as spins in the double dots form a triplet state that blocks the current flow (Fig. 14(b) right) [6]. In the PSB regime, we can readout the spin state, as higher current means one of the triplet spins has been flipped. With SOC, we control the spin by applying microwave (MW) burst to the quantum dot gate. When the MW frequency equals the spin Zeeman splitting by the magnet field, spins can be flipped, as shown in Fig. 15 [6]. Finally, we demonstrate coherent spin control with Rabi oscillations [9]. By sweeping the MW burst time, the spin oscillates up and down with a Rabi frequency ~ 18 MHz, as shown in Fig. 16(a). We further show the quadratic dependency of Rabi frequency with MW power, as the spin control speed is proportional to the MW amplitude (Fig. 16(b)).

## V. SUMMARY & FUTURE WORK

Scaling up the qubit number and further improving the qubit performance are the main research focuses for silicon quantum dot qubit systems. We address them by developing a 300 mm process for qubit device integrations. Through this process, we incorporate different device designs, achieve highly uniform gates with $3\sigma$ ~ 5 nm, and demonstrate almost 100% device yield for 30 nm gate spacings. We further verify the process flow and device designs by cryogenic measurements, demonstrating well-defined single charges, highly tunable tunnel couplings, and coherent spin operations. Future efforts involve incorporating more materials and structures into the qubit devices in the fab process and improving cryogenic characterization throughputs by low temperature multiplexing and clever device designs. Through this learning cycle, knowledge on relevant material properties can be effectively collected, which will pave the way for large-scale silicon spin quantum processors.

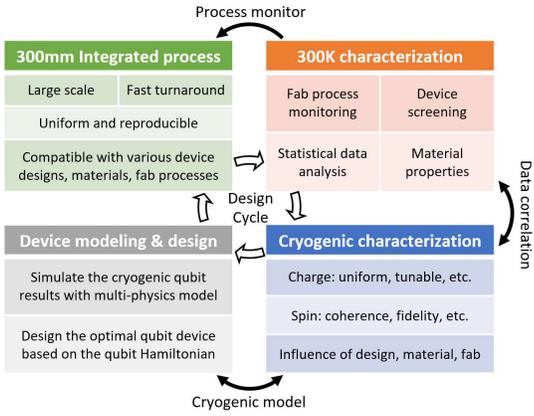

Fig. 1. Flowchart representation of the design cycle. The cycle is designed towards a detailed understanding of material properties at cryogenic temperatures for qubit up-scaling. It relies on close links between integration, characterization, and modelling. The key requirements for each part are summarized in the sub-tables.

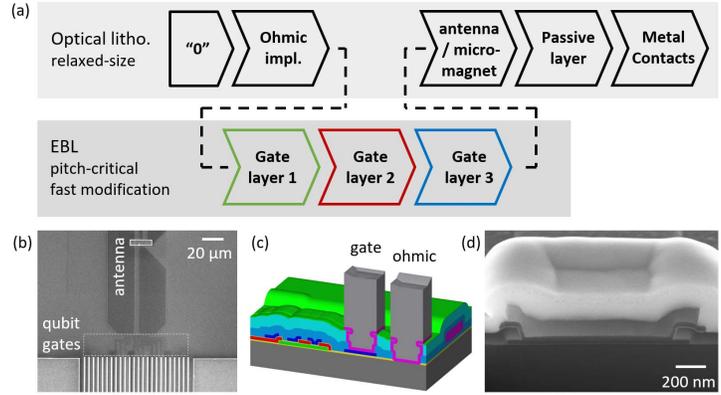

Fig. 2. (a) Schematic of the hybrid 300 mm fab integration flow for qubit devices. Qubit gates are patterned by EBL and surrounding relaxed-size structures are patterned by optical lithography. (b) SEM image of the qubit gate region. (c) Schematic of the BEOL module and contacting vias stopping on an EBL gate and ohmic. (d) Cross-section SEM image of the via contact between the optical lithography metal and the EBL qubit gate.

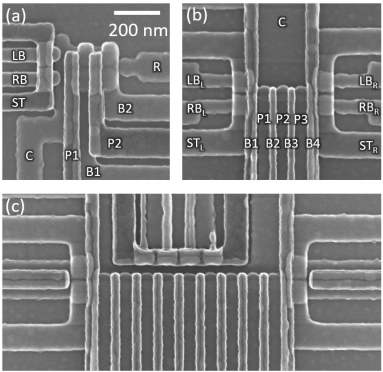

Fig. 3. SEM images of EBL qubit gates designed for (a) double quantum dots under P1 and P2 with a charge sensor dot between LB and RB; (b) triple quantum dots under P1, P2, and P3; and (c) a 1D quantum dot array.

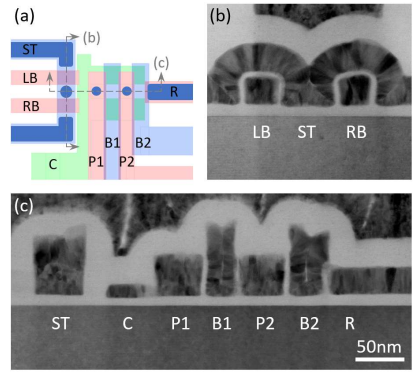

Fig. 4. (a) Schematic of a qubit device similar to Fig. 3(a). Gate layers 1 / 2 / 3 are shown as green / red / blue. Dark blue denotes charge carriers. (b) and (c) are the TEM cross-section at the charge sensor dot region and the double dot region, respectively.

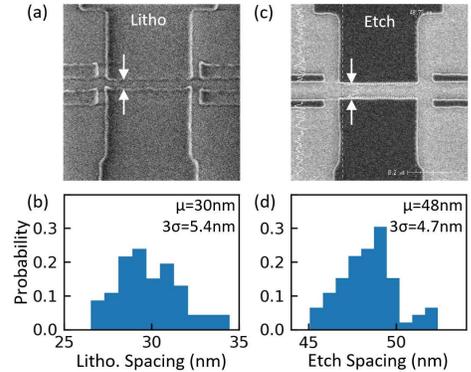

Fig. 5. CDSEM measurement on the confinement gate gap. (a) and (c) show resist after EBL and the gate after etching, respectively. (b) and (d) show the histogram of over 50 CDSEM across the 300 mm wafer on the structures in (a) and (c), respectively.

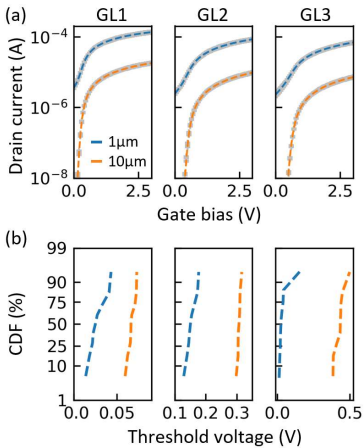

Fig. 6. (a) $I_dV_g$ curves for metrology nMOS transistors with channel length L = 1 and 10 μm on different gate layers. (b) CDF of the threshold voltages. For long channel transistors, threshold increases with gate layers, while this trend is less clear for short channel transistors.

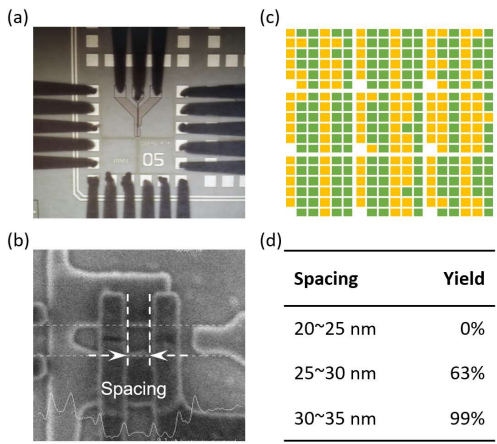

Fig. 7. (a) A customized probe card aligned to the bond pads of a qubit device. (b) The gate structure undergoes the probe card testing. Gate spacings are varied across the wafer, and inter-gate shorts are measured. (c) Probe card result matrix. Each box denotes a device. Spacings are changed column-wise. From left, the spacings are 20~25 nm, 25~30 nm, 30~35 nm, and repeat. Green (yellow) means no short (short). (d) The yield rate table.

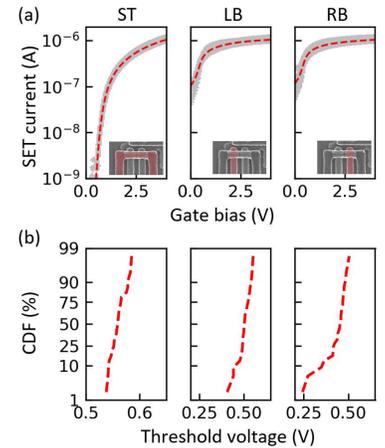

Fig. 8. (a) $I_dV_g$ curves for the SET gates. Inserts show the structures and sweeping gates are false colored red. The non-sweeping gates are kept at 3.5 V. (b) CDF of the threshold voltages, which are defined by linear fitting to the sub-threshold regions to increase the sensitivity to gate topographies.

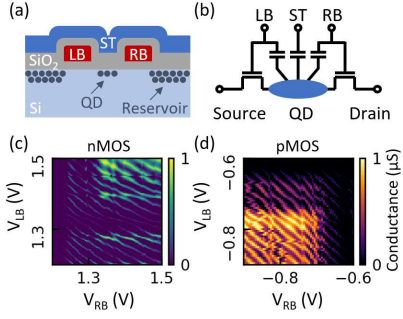

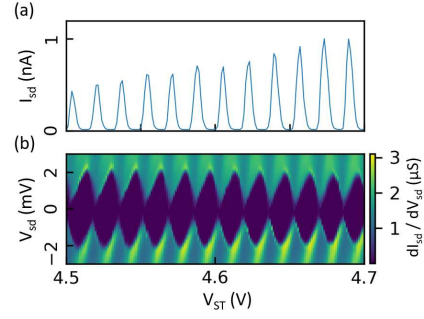

Fig. 9. (a) Schematic cross-section of a single quantum dot defined by similar gates as Fig. 4(b). (b) Equivalent circuit of a single dot. (c) and (d) show the conductance through the single dot by sweeping LB and RB for a nMOS and pMOS dot, respectively. The ~ 45° high conductance lines indicate the dot is equally coupled to LB and RB.

Fig. 10. (a) Coulomb oscillations of a SET measured at $V_{sd} = 1$ mV. (b) Coulomb diamonds of the SET. The differential conductance is measured by standard lock-in technique with an ac excitation $dV_{sd} = 100$ μV.

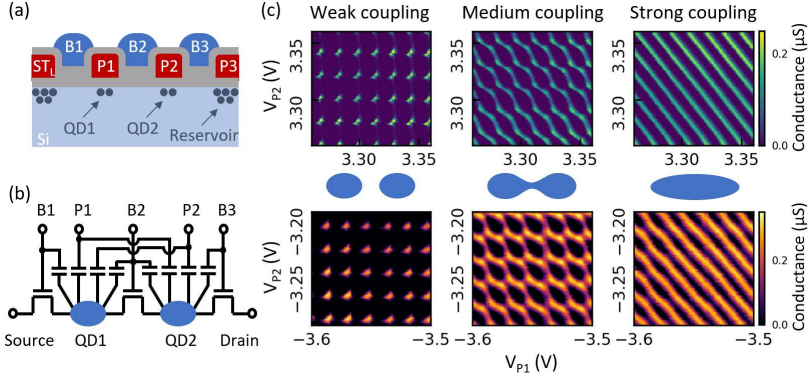

Fig. 11. (a) Schematic cross-section of a double dot defined by similar gates as Fig. 3(b). (b) Equivalent circuit of a double dot. (c) Tunable tunnel coupling between the double dot for nMOS (upper) and pMOS (lower) devices. The plunger gates P1 and P2 control the potential of dot 1 and 2, respectively. B2 controls the inter-dot coupling. In the weak coupling regime, dot 1 and 2 are separated. Conductance maps show isolated points when energy levels between dots are aligned. In the medium coupling regime, levels of double dots begin to hybridize. Conductance maps show honeycomb patterns. In the strong coupling regime, two dots merge into one. Conductance maps show single dot features similar to Fig. 9 (c-d).

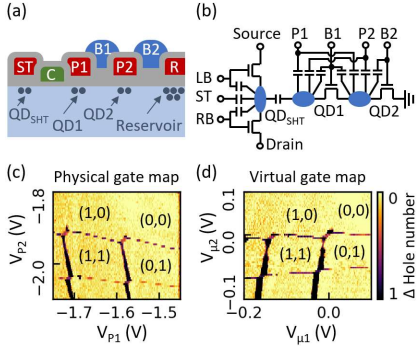

Fig. 12. (a) Schematic cross-section of a double dot with a charge sensor. The gate structure is similar to Fig. 4(c). (b) Equivalent circuit. (c) Charge sensing map on a pMOS device. Brackets denote the hole numbers in dot 1 and 2. (d) Remap of (c) in the virtual gate space. Crosstalk between P1 and P2 is removed.

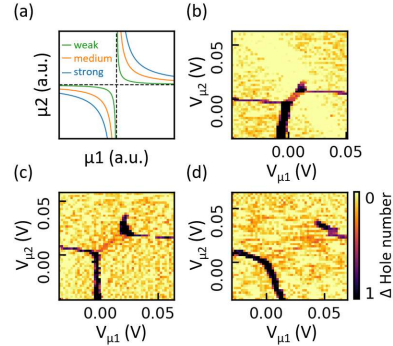

Fig. 13. (a) Simulated energy diagram of inter-dot transitions with different coupling strength. (b-d) are zoom in maps in Fig. 12(d) at the (1,0) - (0,1) transition region. By tuning the B1 bias, the inter-dot tunnel coupling can be changed from weak to strong with B1, resembling the patterns in (a).

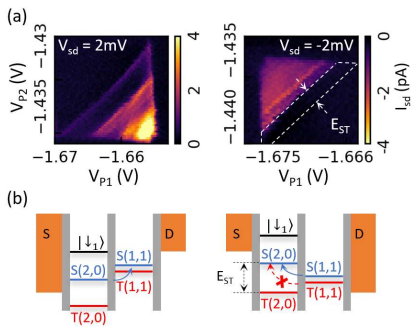

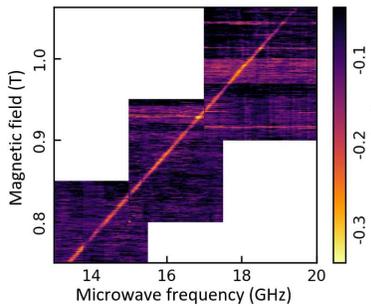

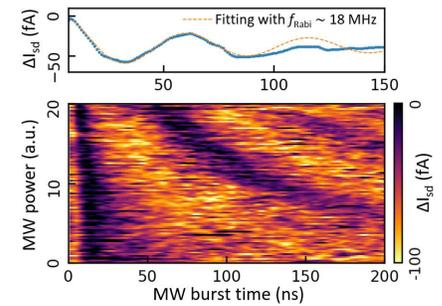

Fig. 14. (a) Bias triangles of a pMOS double dot system in weak coupling regime like that of Fig. 11 but with higher $|V_{sd}|$. For $V_{sd} = -2$ mV, the triangle bottom is missing due to spin Pauli blockade: When there is an unpaired spin in dot 1, the upcoming spin from dot 2 will be blocked if they form a triplet state, as shown in (b) right. For $V_{sd} = 2$ mV, the extra spin enters dot 1 from the source reservoir, and hence a singlet state can always be loaded, as shown in (b) left.

Fig. 15. Double dot current in the PSB region as a function of external magnetic field and the frequency of MW burst applied on the double dot gate. With spin-orbit interaction, the spin of holes can be flipped when the MW photon frequency equals the spin Zeeman energy. Spin flipping lifts the PSB, increasing the current through the double dot. The high current line remarks the linear relation between magnetic field and MW frequency for on-resonance spin flip.

Fig. 16. (a) Coherent spin control with Rabi oscillations. When the MW frequency is tuned to be on-resonance with the spin Zeeman energy, the spin direction can be controlled coherently by changing the MW burst time. The oscillation of spin direction is reflected by the oscillation of current in the PSB regime. (b) Power dependency of Rabi oscillations. The Rabi frequency depends on the MW amplitude. Therefore, the peaks or dips of the Rabi oscillations shift quadratically when stepping the MW power.